\documentclass[11pt,preprintnumbers,titlepage,nofootinbib]{revtex4-2}
\usepackage[latin9]{inputenc}
\usepackage{amsmath}
\usepackage{amssymb}
\usepackage{graphicx}
\usepackage[unicode=true,
 bookmarks=false,
 breaklinks=false,pdfborder={0 0 1},backref=section,colorlinks=false]
 {hyperref}
\hypersetup{
 colorlinks,linkcolor=blue,citecolor=blue,urlcolor=blue}

\makeatletter

\newcommand{\lyxdot}{.}

\usepackage{wasysym}
\usepackage{array}
\usepackage{multirow}
\usepackage{wasysym}
\usepackage{wasysym}
\usepackage{amsfonts}
\usepackage{subfigure}
\usepackage{cleveref}
\usepackage{listings}
\usepackage{xcolor}
\usepackage{array}
\usepackage{url}
\usepackage{color}
\usepackage{subfigure}
\@ifundefined{textcolor}{}{
\definecolor{BLACK}{gray}{0}
\definecolor{WHITE}{gray}{1}
\definecolor{RED}{rgb}{1,0,0}
\definecolor{GREEN}{rgb}{0,1,0}
\definecolor{BLUE}{rgb}{0,0,1}
\definecolor{CYAN}{cmyk}{1,0,0,0}
\definecolor{MAGENTA}{cmyk}{0,1,0,0}
\definecolor{YELLOW}{cmyk}{0,0,1,0}
\definecolor{ballblue}{rgb}{0.13, 0.67, 0.8}
\definecolor{bleudefrance}{rgb}{0.19, 0.55, 0.91}
\definecolor{blue(ncs)}{rgb}{0.0, 0.53, 0.74}
\definecolor{darkpastelgreen}{rgb}{0.01, 0.75, 0.24}
\definecolor{darkspringgreen}{rgb}{0.09, 0.45, 0.27}
\definecolor{denim}{rgb}{0.08, 0.38, 0.74}
\definecolor{electricviolet}{rgb}{0.56, 0.0, 1.0}
}

\makeatother

\begin{document}
\preprint{CTP-SCU/2023037}
\title{Observations of Orbiting Hot Spots around Naked Singularities}
\author{Yiqian Chen$^{a}$}
\email{yqchen@stu.scu.edu.cn}

\author{Peng Wang$^{a}$}
\email{pengw@scu.edu.cn}

\author{Houwen Wu$^{a,b}$}
\email{hw598@damtp.cam.ac.uk}

\author{Haitang Yang$^{a}$}
\email{hyanga@scu.edu.cn}

\affiliation{$^{a}$Center for Theoretical Physics, College of Physics, Sichuan
University, Chengdu, 610064, China}
\affiliation{$^{b}$Department of Applied Mathematics and Theoretical Physics,
University of Cambridge, Wilberforce Road, Cambridge, CB3 0WA, UK}
\begin{abstract}
Recently, it has been reported that photons can traverse naked singularities
in the Janis-Newman-Winicour and Born-Infeld spacetimes when these
singularities are appropriately regularized. In this paper, we investigate
observational signatures of hot spots orbiting these naked singularities,
with a focus on discerning them from black holes. In contrast to Schwarzschild
black holes, we unveil the presence of multiple additional image tracks
within critical curves in time integrated images capturing a complete
orbit of hot spots. Moreover, these new images manifest as a more
pronounced second-highest peak in temporal magnitudes when observed
at low inclinations.
\end{abstract}
\maketitle
\tableofcontents{}

{}

{}

\section{Introduction}

\label{sec:Introduction}

The recent remarkable advancement in high angular resolution achieved
by the Event Horizon Telescope (EHT) collaboration has ushered in
a new era for the study of gravitational lensing within the context
of strong gravitational fields \cite{Akiyama:2019cqa,Akiyama:2019brx,Akiyama:2019sww,Akiyama:2019bqs,Akiyama:2019fyp,Akiyama:2019eap,Akiyama:2021qum,Akiyama:2021tfw,EventHorizonTelescope:2022xnr,EventHorizonTelescope:2022vjs,EventHorizonTelescope:2022wok,EventHorizonTelescope:2022exc,EventHorizonTelescope:2022urf,EventHorizonTelescope:2022xqj}.
This development has kindled a profound interest in the examination
of black hole images that are illuminated by the accreting plasma.
The extraordinary black hole images captured by the EHT have opened
the possibility to directly test sophisticated theoretical models,
such as General Relativistic Magnetohydrodynamical (GRMHD) numerical
simulations, against observations. Given the substantial computational
resources required for GRMHD simulations, researchers often resort
to simplified accretion models that, while computationally more tractable,
sufficiently capture the fundamental characteristics of black hole
images \cite{Shaikh:2018lcc,Narayan:2019imo,Zeng:2020dco,Zeng:2020vsj,Saurabh:2020zqg,Qin:2020xzu,Luminet:1979nyg,Beckwith:2004ae,Gralla:2019xty,Dokuchaev:2019pcx,Peng:2020wun,He:2021htq,Eichhorn:2021iwq,Li:2021riw,Gan:2021pwu,Gan:2021xdl}.

A noteworthy characteristic of these images is the presence of a shadow
region, encircled by a luminous ring. This distinctive feature arises
from strong gravitational lensing effects in the vicinity of unstable
bound photon orbits \cite{Synge:1966okc,Bardeen:1972fi,Bardeen:1973tla,Virbhadra:1999nm,Claudel:2000yi,Virbhadra:2008ws,Bozza:2009yw,Virbhadra:2022iiy}.
The black hole shadow, as observed by the EHT, is anticipated to carry
vital information about the spacetime geometry surrounding the black
hole. Remarkably, its features closely align with the predictions
based on the Kerr black hole model. Nevertheless, it is important
to acknowledge that uncertainties related to the black hole's mass-to-distance
ratio and potential systematic errors within the EHT observations
introduce some degree of ambiguity within the bounds of observational
uncertainty, allowing for the possibility of alternatives to Kerr
black holes. Furthermore, recent discoveries of horizonless ultra-compact
objects that exhibit photon spheres have added another layer of complexity
to the scenario, effectively mimicking black holes in various observational
simulations \cite{Schmidt:2008hc,Guzik:2009cm,Liao:2015uzb,Goulart:2017iko,Nascimento:2020ime,Islam:2021ful,Tsukamoto:2021caq,Junior:2021svb,Olmo:2021piq,Ghosh:2022mka}.

Among ultra-compact objects, naked singularities have attracted considerable
attention. Although the cosmic censorship conjecture prohibits the
formation of naked singularities, they can arise through the gravitational
collapse of massive objects under specific initial conditions \cite{Shapiro:1991zza,Joshi:1993zg,Harada:1998cq,Joshi:2001xi,Goswami:2006ph,Banerjee:2017njk,Bhattacharya:2017chr}.
\ The presence of photon spheres allows naked singularities to effectively
mimic the optical characteristics of their black hole counterparts,
instigating inquiries into the distinctive observational imprints
attributable to naked singularities \cite{Virbhadra:2002ju,Virbhadra:2007kw,Gyulchev:2008ff,Sahu:2012er,Shaikh:2019itn,Paul:2020ufc,Zhdanov:2019ozq,Stashko:2021lad,Stashko:2021het,Tsukamoto:2021fsz,Wang:2023jop,Chen:2023trn}.
Interestingly, in certain naked singularity spacetimes, it has been
established that photons can both approach and depart from the singularities
in finite coordinate time intervals \cite{Shaikh:2018lcc,Chen:2023trn,Chen:2023uuy}.
In such spacetimes, images of naked singularities captured by distant
observers critically depend on the intrinsic nature of these singularities---a
facet that demands a deeper exploration through a quantum gravity
framework. However, the absence of a definitive theory of quantum
gravity poses formidable challenges when it comes to investigating
the behavior of photons in the proximity of singularities. Consequently,
researchers frequently resort to effective models for singularity
regularization, thus enabling the study of null geodesics near these
points. One such approach involves the incorporation of higher-order
curvature terms, such as the complete $\alpha^{\prime}$ corrections
of string theory \cite{Wang:2019kez,Wang:2019dcj,Ying:2021xse}.

Recently, our investigations have centered on the phenomenon of gravitational
lensing applied to distant light sources within the context of Janis-Newman-Winicour
(JNW) and Born-Infeld singularities. Our findings have revealed that
photons entering the photon spheres ultimately converge toward the
singularities in a finite coordinate time \cite{Chen:2023trn,Chen:2023uuy}.
When these singularities are subjected to regularization through the
introduction of a regular core, it becomes possible for these photons
to traverse the now regularized singularity. This traversal results
in the emergence of new images occurring within critical curves. In
this present study, our focus is toward observational properties of
JNW and Born-Infeld singularities when illuminated by localized and
isotropically emitting sources, referred to as ``hot spots.''\ 

In certain GRMHD simulations and semi-analytic models, the occurrence
of magnetic reconnection and flux eruptions yields the formation of
hot spots encircling supermassive black holes that host a magnetized
accretion disk \cite{Dexter:2020cuv,Scepi:2021xgs,ElMellah:2021tjo}.
Notably, these hot spots have been recurrently observed within the
vicinity of Sgr A{*} \cite{Witzel:2020yrp,Michail:2021pgd,GRAVITY:2021hxs}.
Furthermore, a noteworthy instance involves the detection of an orbiting
hot spot within the unresolved light curve data obtained at the observing
frequency of the EHT \cite{Wielgus:2022heh}. Due to their origin
from a compact region proximate to the innermost stable circular orbit
(ISCO), these hot spots represent a promising tool for the examination
of central objects in the strong gravity regime \cite{abuter2018detection,Wielgus:2022heh}.

The subsequent sections of this paper are structured as follows: In
Section \ref{sec:SandG}, we briefly review the JNW and Born-Infeld
singularities, along with a discussion of geodesic motion within these
spacetimes. Section \ref{sec:Observational-of-orbiting} is devoted
to the hot spot model, followed by an examination time integrated
images, temporal fluxes and centroids. Finally, Section \ref{sec:CONCLUSIONS}
presents our conclusions. We adopt the convention $G=c=1$ throughout
the paper.

\section{Spacetime and Geodesics}

\label{sec:SandG}

In this section, we provide a concise overview of both JNW and Born-Infeld
singularities, while also examining the geodesic motion within these
spacetimes. For a spherically symmetric and static spacetime governed
by the metric 
\begin{equation}
ds^{2}=-f\left(r\right)dt^{2}+\frac{1}{h\left(r\right)}dr^{2}+R\left(r\right)\left(d\theta^{2}+\sin^{2}\theta d\varphi^{2}\right),\label{eq:metric}
\end{equation}
the trajectory of a test particle with four-momentum $p^{\mu}$ is
determined by the geodesic equations
\begin{equation}
\frac{dx^{\mu}}{d\lambda}=p^{\mu},\quad\frac{dp^{\mu}}{d\lambda}=-\Gamma_{\rho\sigma}^{\mu}p^{\rho}p^{\sigma}.\label{eq:geoeq}
\end{equation}
Here, $\lambda$ is the affine parameter, and $\Gamma_{\rho\sigma}^{\mu}$
indicates the Christoffel symbol. These geodesics are fully characterized
by three conserved quantities, 
\begin{equation}
E=-p_{t},\quad L_{z}=p_{\varphi},\quad L^{2}=p_{\theta}^{2}+L_{z}^{2}\csc^{2}\theta.
\end{equation}

In the context of massless particles, the conserved quantities $E$,
$L_{z}$ and $L$ represent the total energy, the angular momentum
parallel to the axis of symmetry and the total angular momentum, respectively.
Additionally, the Hamiltonian constraint $\mathcal{H}\equiv g_{\mu\nu}p^{\mu}p^{\nu}/2=0$
yields the radial component of the null geodesic equations as
\begin{equation}
\dot{r}^{2}+V_{\text{eff}}\left(r\right)=0,
\end{equation}
where the dot signifies differentiation with respect to an affine
parameter $\lambda$, and the introduced effective potential is given
by 
\begin{equation}
V_{\text{eff}}\left(r\right)=h\left(r\right)\left[\frac{L^{2}}{R\left(r\right)}-\frac{E^{2}}{f\left(r\right)}\right].
\end{equation}
A circular null geodesic occurs at an extremum of the effective potential
$V_{\text{eff}}(r)$, and the radius $r_{c}$ of this geodesic is
determined by the conditions 
\begin{equation}
V_{\text{eff}}\left(r_{c}\right)=0,\text{ }V_{\text{eff}}^{\prime}\left(r_{c}\right)=0.
\end{equation}
Furthermore, local maxima and minima of the effective potential correspond
to unstable and stable circular null geodesics, respectively. These
unstable and stable circular null geodesics constitute a photon sphere
and an anti-photon sphere, respectively.

For massive particles, $E$, $L_{z}$ and $L$ represent the total
energy per unit mass, the angular momentum per unit mass parallel
to the axis of symmetry and the total angular momentum per unit mass,
respectively, when the affine parameter $\lambda$ is chosen as the
proper time per unit mass. Similarly, the Hamiltonian constraint $\mathcal{H}=-1/2$
leads to the effective potential 
\begin{equation}
V_{\text{eff}}\left(r\right)=h\left(r\right)\left[\frac{L^{2}}{R\left(r\right)}-\frac{E^{2}}{f\left(r\right)}+1\right].
\end{equation}
Consequently, the ISCO at $r=r_{\text{ISCO}}$ is determined by the
conditions
\begin{equation}
V_{\text{eff}}\left(r_{\text{ISCO}}\right)=0\text{, }V_{\text{eff}}^{\prime}\left(r_{\text{ISCO}}\right)=0\text{, }V_{\text{eff}}^{^{\prime\prime}}\left(r_{\text{ISCO}}\right)=0.
\end{equation}

\subsection{JNW Singularity}

The JNW metric provides a static solution within Einstein-massless-scalar-field
models and is expressed in the form \cite{Fisher:1948yn,Janis:1968zz,Wyman:1981bd,Virbhadra:1995iy}
\begin{equation}
ds^{2}=-\left(1-\frac{r_{g}}{r}\right)^{\gamma}dt^{2}+\left(1-\frac{r_{g}}{r}\right)^{-\gamma}dr^{2}+\left(1-\frac{r_{g}}{r}\right)^{1-\gamma}r^{2}\left(d\theta^{2}+\sin^{2}\theta d\varphi^{2}\right).
\end{equation}
Additionally, the scalar field is given by
\begin{equation}
\Phi=\frac{q}{r_{g}}\ln\left(1-\frac{r_{g}}{r}\right),
\end{equation}
where $q$ denotes the scalar charge. The JNW metric is characterized
by two parameters, $\gamma$ and $r_{g}$, which are related to the
ADM mass $M$ and the scalar charge $q$ according to \cite{Janis:1968zz},
\begin{equation}
\gamma=\frac{2M}{r_{g}},\text{ }r_{g}=2\sqrt{M^{2}+q^{2}}.
\end{equation}
When $\gamma=1$, the JNW metric describes Schwarzschild black holes
with no scalar charge. For $0.5<\gamma<1$, the JNW metric represents
weakly naked singularity solutions with a non-trivial scalar field
profile. In this case, a naked curvature singularity arises at $r=r_{g}$,
and a photon sphere exists at $r_{ps}=r_{g}\left(1+2\gamma\right)/2$.
However, the photon sphere disappears when $0\leq\gamma<0.5$, leading
to distinct light propagation behaviors. Given that a spacetime featuring
photon spheres can mimic black hole observations, this paper primarily
focuses on the JNW metric with $0.5<\gamma<1$.

As shown in \cite{Chen:2023uuy}, null geodesics in the vicinity of
the singularity can be expressed as
\begin{align}
t\left(\lambda\right) & =t_{0}\pm_{r}\frac{E^{1-\gamma}{}r_{g}^{\gamma}\left\vert \lambda\right\vert ^{1-\gamma}}{1-\gamma}+\mathcal{O}\left(\left\vert \lambda\right\vert {}^{1-\gamma}\right),\nonumber \\
r\left(\lambda\right) & =r_{g}\pm_{r}E\lambda+\mathcal{O}\left(\left\vert \lambda\right\vert ^{\frac{1}{2-2\gamma}}{}\right),\nonumber \\
\theta\left(\lambda\right) & =\theta_{0}\pm_{\theta}\sqrt{L^{2}-L_{z}^{2}\csc^{2}\theta_{0}}\frac{E^{\gamma-1}\left\vert \lambda\right\vert ^{\gamma}}{\gamma r_{g}^{1+\lambda}}+\mathcal{O}\left(\left\vert \lambda\right\vert {}^{\gamma}\right),\label{eq:JNWgeo}\\
\varphi\left(\lambda\right) & =\varphi_{0}+\frac{L_{z}E^{\gamma-1}\csc^{2}\theta_{0}\left\vert \lambda\right\vert ^{\gamma}}{\gamma r_{g}^{\gamma+1}}+\mathcal{O}\left(\left\vert \lambda\right\vert ^{\gamma}\right),\nonumber 
\end{align}
where $t_{0}$, $\theta_{0}$ and $\varphi_{0}$ are the integration
constants, and we assume $r\left(0\right)=r_{g}$. It shows the existence
of two classes of light rays: radially outgoing and ingoing light
rays, denoted as $+_{r}$ and $-_{r}$, respectively. To simplify,
we adopt $\lambda<0$ for ingoing light rays and $\lambda>0$ for
outgoing ones. As the affine parameter $\lambda$ approaches $0$
from the right and left, respectively, both outgoing and ingoing light
rays converge toward the singularity.

Interestingly, as indicated by eqn. $\left(\ref{eq:JNWgeo}\right)$,
it becomes apparent that photons originating from distant sources
can reach the singularity in a finite coordinate time, and conversely,
photons escaping from the singularity only require a finite coordinate
time to reach distant observers. This aspect is of particular interest
since, from the perspective of distant observers, whose proper time
is approximately the coordinate time, the destiny of photons at the
singularity significantly influences the observable characteristics
of the JNW naked singularity. In the quest to examine the behavior
of photons in close proximity to the singularity, researchers often
resort to effective models to regularize the singularity. Specifically,
the singularity can be regularized with an infinitesimally small regular
core, as outlined in \cite{Chen:2023uuy}. In this regularized singularity
spacetime, light rays, upon entering the photon sphere, transverse
the regular core and can be accurately approximated by a composite
of the ingoing and outgoing branches given in eqn. $\left(\ref{eq:JNWgeo}\right)$.
Furthermore, the connection between these two branches is given by
\begin{equation}
\theta(0_{-})=\pi-\theta(0_{+}),\quad\varphi(0_{-})=\pi+\varphi(0_{+}).\label{eq:cc}
\end{equation}
In short, the condition $\left(\ref{eq:cc}\right)$ and the conservation
of $E$, $L_{z}$ and $L$ determine the corresponding outgoing branch
for a given ingoing branch.

\subsection{Born-Infeld Singularity}

The Born-Infeld metric is a set of spherically symmetric and static
solutions arising from an Einstein gravity model coupled with a Born-Infeld
electromagnetic field, which is presented in \cite{Dey:2004yt,Cai:2004eh,Guo:2022ghl}.
This metric can be expressed as 
\begin{equation}
ds^{2}=-f_{\text{BI}}\left(r\right)dt^{2}+\frac{dr^{2}}{f_{\text{BI}}\left(r\right)}+r^{2}\left(d\theta^{2}+\sin^{2}\theta d\varphi^{2}\right)\text{,}\label{eq:NLEDBH}
\end{equation}
where 
\begin{equation}
f_{\text{BI}}\left(r\right)=1-\frac{2M}{r}-\frac{2\left(Q^{2}+P^{2}\right)}{3\sqrt{r^{4}+a\left(Q^{2}+P^{2}\right)}+3r^{2}}+\frac{4\left(Q^{2}+P^{2}\right)}{3r^{2}}\text{ }_{2}F_{1}\left(\frac{1}{4},\frac{1}{2},\frac{5}{4};-\frac{a\left(Q^{2}+P^{2}\right)}{r^{4}}\right).
\end{equation}
Here, the parameter $a$ is related to the string tension $\alpha^{\prime}$
as $a=\left(2\pi\alpha^{\prime}\right)^{2}$ while the black hole's
mass, electrical charge, and magnetic charge are denoted as $M$,
$Q$ and $P$, respectively. The hypergeometric function $_{2}F_{1}\left(a,b,c;x\right)$
is employed. Depending on these parameters, the Born-Infeld metric
can describe either a black hole or a naked singularity. The domain
of existence for naked singularities within the parameter space $a/M^{2}$-$\sqrt{Q^{2}+P^{2}}/M$
has been illustrated in \cite{Guo:2022ghl}.

Moreover, the nonlinearity of Born-Infeld electrodynamics introduces
self-interaction of the electromagnetic field. Consequently, photons
follow null geodesics in an effective metric with the metric functions
\cite{Novello:1999pg} 
\begin{align}
f(r) & =\frac{(aP^{2}+r^{4})^{2}}{r^{2}\left[a\left(Q^{2}+P^{2}\right)+r^{4}\right]^{3/2}}f_{\text{BI}}(r),\nonumber \\
h(r) & =\frac{r^{2}\left[a\left(Q^{2}+P^{2}\right)+r^{4}\right]^{3/2}}{(aP^{2}+r^{4})^{2}}f_{\text{BI}}(r)\text{,}\label{eq:geff-BI}\\
R(r) & =\frac{(aP^{2}+r^{4})^{2}}{r^{4}\sqrt{a\left(Q^{2}+P^{2}\right)+r^{4}}}.\nonumber 
\end{align}
Furthermore, our investigation reveals the behavior of null geodesics
in this effective metric near the singularity as detailed in \cite{Chen:2023trn},
\begin{align}
t\left(\lambda\right) & =t_{0}\pm_{r}\frac{3\sqrt{\pi}a^{5/4}\left(Q^{2}+P^{2}\right)\lambda^{-2}}{8\Gamma\left(1/4\right)\Gamma\left(5/4\right)\left(Q^{2}+P^{2}\right)^{3/2}E^{2}-12\sqrt{\pi}a^{1/4}E^{2}M}+\mathcal{O}\left(\left\vert \lambda\right\vert {}^{-3}\right),\nonumber \\
r\left(\lambda\right) & =\pm_{r}\frac{\sqrt{a\left(Q^{2}+P^{2}\right)}}{E}\lambda^{-1}+\mathcal{O}\left(\left\vert \lambda\right\vert {}^{-2}\right),\nonumber \\
\theta\left(\lambda\right) & =\theta_{0}\pm_{\theta}\frac{\sqrt{a\left(Q^{2}+P^{2}\right)}\sqrt{L^{2}-L_{z}^{2}\csc^{2}\theta_{0}}}{3E}\lambda^{-3}+\mathcal{O}\left(\left\vert \lambda\right\vert {}^{-4}\right),\\
\varphi\left(\lambda\right) & =\varphi_{0}+\frac{\sqrt{a\left(Q^{2}+P^{2}\right)}L_{z}\csc^{2}\theta_{0}}{3E^{4}}\lambda^{-3}+\mathcal{O}\left(\left\vert \lambda\right\vert {}^{-4}\right).\nonumber 
\end{align}
Here, the upper and lower signs of $\pm_{r}$ correspond to the radially
outgoing and ingoing branches, respectively. Additionally, we adopt
$\lambda>0$ for the outgoing branch and and $\lambda<0$ for the
ingoing branch, respectively. It is worth emphasizing that\ the affine
parameter approaches $\pm\infty$ when the light ray approaches the
singularity. Much like the situation with JNW singularities, photons
characterized by suitably small impact parameters can traverse the
regularized singularity in a finite coordinate time. The trajectories
of these photons are effectively approximated by the outgoing and
ingoing branches, and their connection is established through 
\begin{equation}
\theta(-\infty)=\pi-\theta(\infty)\quad\text{and }\varphi(-\infty)=\pi+\varphi(\infty).
\end{equation}

Remarkably, it has been demonstrated that Born-Infeld naked singularity
solutions can possess two photon spheres and one anti-photon sphere
in the effective metric. In \cite{Chen:2023trn}, the parameter regions
where two photon spheres with distinct sizes exist are presented in
the $a/M^{2}$-$\sqrt{P^{2}+Q^{2}}/M$ parameter space. Particularly,
we focus on scenarios where the potential peak at the inner photon
sphere is higher than that of the outer sphere. In such instances,
both photon spheres can contribute to determining optical appearances
of Born-Infeld naked singularities.

\section{Observation of Hot Spot}

\label{sec:Observational-of-orbiting}

This section is dedicated to the examination of observational attributes
exhibited by a hot spot encircling both JNW and Born-Infeld naked
singularities. More precisely, we model the hot spot as an isotropically
emitting sphere. Furthermore, this sphere's center revolves around
the central object at a distinct radius $r_{e}$ on the equatorial
plane, propelled by the 4-velocity 
\begin{equation}
v_{e}^{\mu}=\left(\frac{E}{f\left(r_{e}\right)},0,0,\frac{L}{R\left(r_{e}\right)}\right),
\end{equation}
where $E$ and $L$ are given by 
\[
E=\sqrt{\frac{R^{\prime}\left(r_{e}\right)f^{2}\left(r_{e}\right)}{f\left(r_{e}\right)R^{\prime}\left(r_{e}\right)-f^{\prime}\left(r_{e}\right)R\left(r_{e}\right)}},\quad L=\sqrt{\frac{f^{\prime}\left(r_{e}\right)R^{2}\left(r_{e}\right)}{f\left(r_{e}\right)R^{\prime}\left(r_{e}\right)-f^{\prime}\left(r_{e}\right)R\left(r_{e}\right)}}.
\]
Consequently, the corresponding angular velocity and period are $\Omega_{e}=\sqrt{f^{\prime}\left(r_{e}\right)/R^{\prime}\left(r_{e}\right)}$
and $T_{e}=2\pi/\Omega_{e}$, respectively.

To obtain the observed image of the hot spot, we employ the backward
ray-tracing method to compute light rays from the observer to the
hot spot. This involves numerically solving eqn. $\left(\ref{eq:geoeq}\right)$
with appropriate initial conditions at the observer's position, which
is defined by coordinates $\left(t_{o},r_{o},\theta_{o},\varphi_{o}\right)$.
In particular, the initial conditions are determined by considering
the $4$-momentum of photons in the observer's local frame, denoted
as $\left(p^{(t)},p^{(r)},p^{(\theta)},p^{(\varphi)}\right)$. These
local 4-momentum components are related to the 4-momentum $p_{o}^{\mu}=\left.dx^{\mu}/d\lambda\right\vert _{\left(t_{o},r_{o},\theta_{o},\varphi_{o}\right)}$
through the expressions 
\begin{equation}
p^{(t)}=\sqrt{f\left(r_{o}\right)}p_{o}^{t},\quad p^{(r)}=p_{o}^{r}/\sqrt{h\left(r_{o}\right)},\quad p^{(\theta)}=\sqrt{R\left(r_{o}\right)}p_{o}^{\theta},\quad p^{(\varphi)}=\sqrt{R\left(r_{o}\right)}|\sin\theta_{o}|p_{o}^{\varphi}.\label{eq:localP}
\end{equation}
The observation angles $\Theta$ and $\Phi$, defined as per \cite{Cunha:2016bpi},
are given by 
\begin{equation}
\sin\Theta=\frac{p^{(\theta)}}{p},\text{ }\tan\Phi=\frac{p^{(\varphi)}}{p^{(r)}},
\end{equation}
where $p=\sqrt{p^{(r)2}+p^{(\theta)2}+p^{(\varphi)2}}$. For a detailed
explanation of the numerical implementation, interested readers can
refer to \cite{Chen:2023trn}. Within the observer's image plane,
each pixel is associated with Cartesian coordinates $\left(x,y\right)$,
where 
\begin{equation}
x\equiv-r_{o}\Phi,\text{ }y\equiv r_{o}\Theta.
\end{equation}

In our computational framework, the observer's position is $\left(t_{o},r_{o},\theta_{o},\varphi_{o}\right)=\left(t_{o},100M,\theta_{o},\pi\right)$.
The hot spot, with a radius of $0.25M$, orbits counterclockwise along
a circular geodesic at $r_{e}=r_{\text{ISCO}}$. To ensure computational
precision and efficiency, we employ a grid of $1000\times1000$ pixels
for each snapshot and generate $500$ snapshots for a full orbit.
This approach guarantees the production of smoothly evolving images
throughout the period $T_{e}$. At a specific time $t_{k}$, each
pixel within the image plane is assigned an intensity $I_{klm}$,
which collectively forms lensed images of the hot spot. Subsequently,
the analysis focuses on the following image properties \cite{Rosa:2022toh,Rosa:2023qcv},
\begin{itemize}
\item Time integrated image: 
\begin{equation}
\left\langle I\right\rangle _{lm}=\sum\limits _{k}I_{klm}.
\end{equation}
\item Total temporal flux: 
\begin{equation}
F_{k}=\sum\limits _{l}\sum\limits _{m}\Delta\Omega I_{klm},
\end{equation}
where $\Delta\Omega$ corresponds to the solid angle of a pixel.
\item Temporal magnitude: 
\begin{equation}
m_{k}=-2.5\lg\left(\frac{F_{k}}{\min\left(F_{k}\right)}\right).
\end{equation}
\item Temporal centroid: 
\begin{equation}
\overrightarrow{c_{k}}=F_{k}^{-1}\sum\limits _{l}\sum\limits _{m}\Delta\Omega I_{klm}\overrightarrow{r_{lm}},
\end{equation}
where $\overrightarrow{r_{lm}}$ represents the position relative
to the image center. 
\end{itemize}

\subsection{Integrated Images}

\begin{figure}[ptb]
\includegraphics[width=0.49\textwidth]{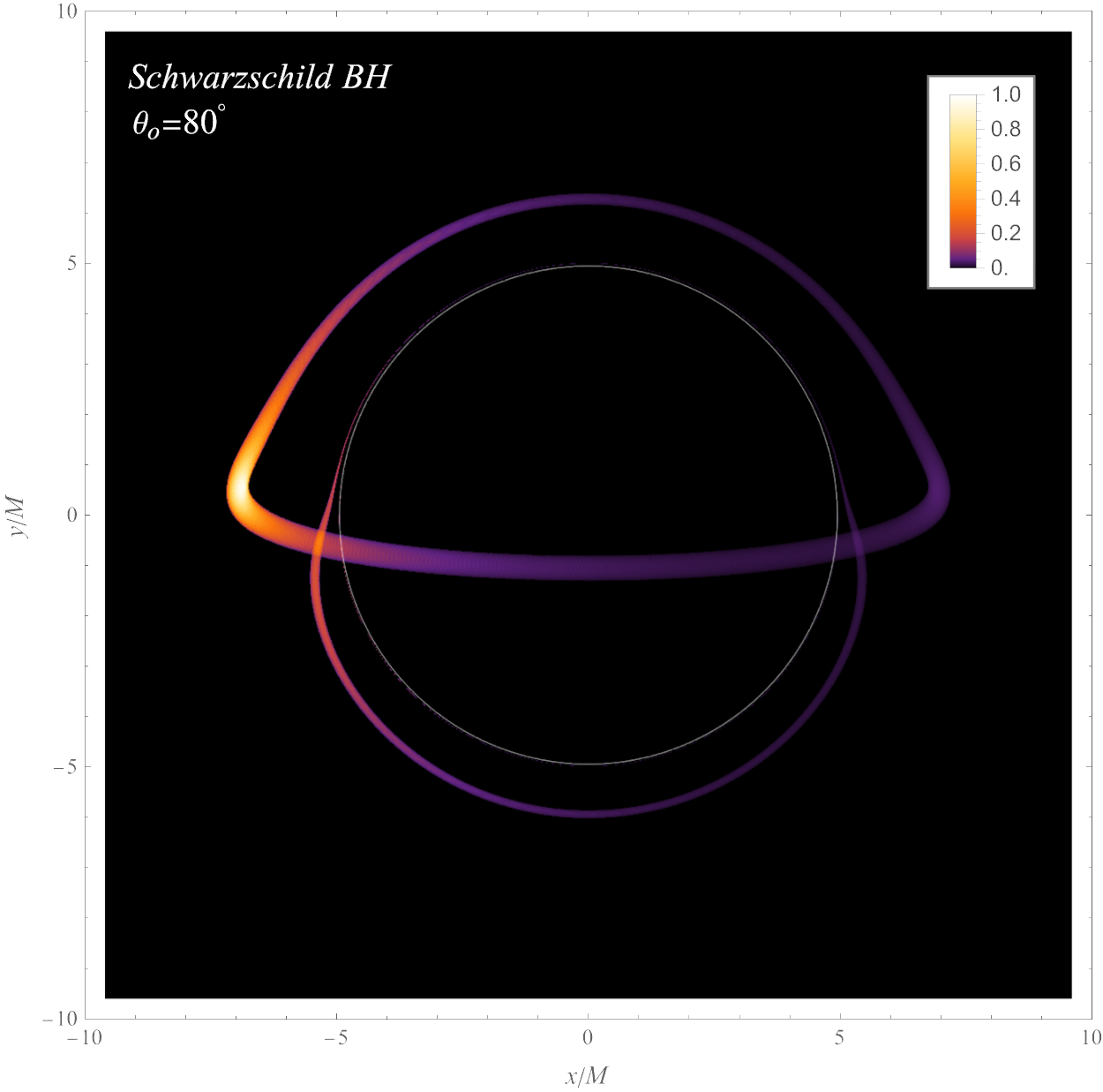}
\includegraphics[width=0.49\textwidth]{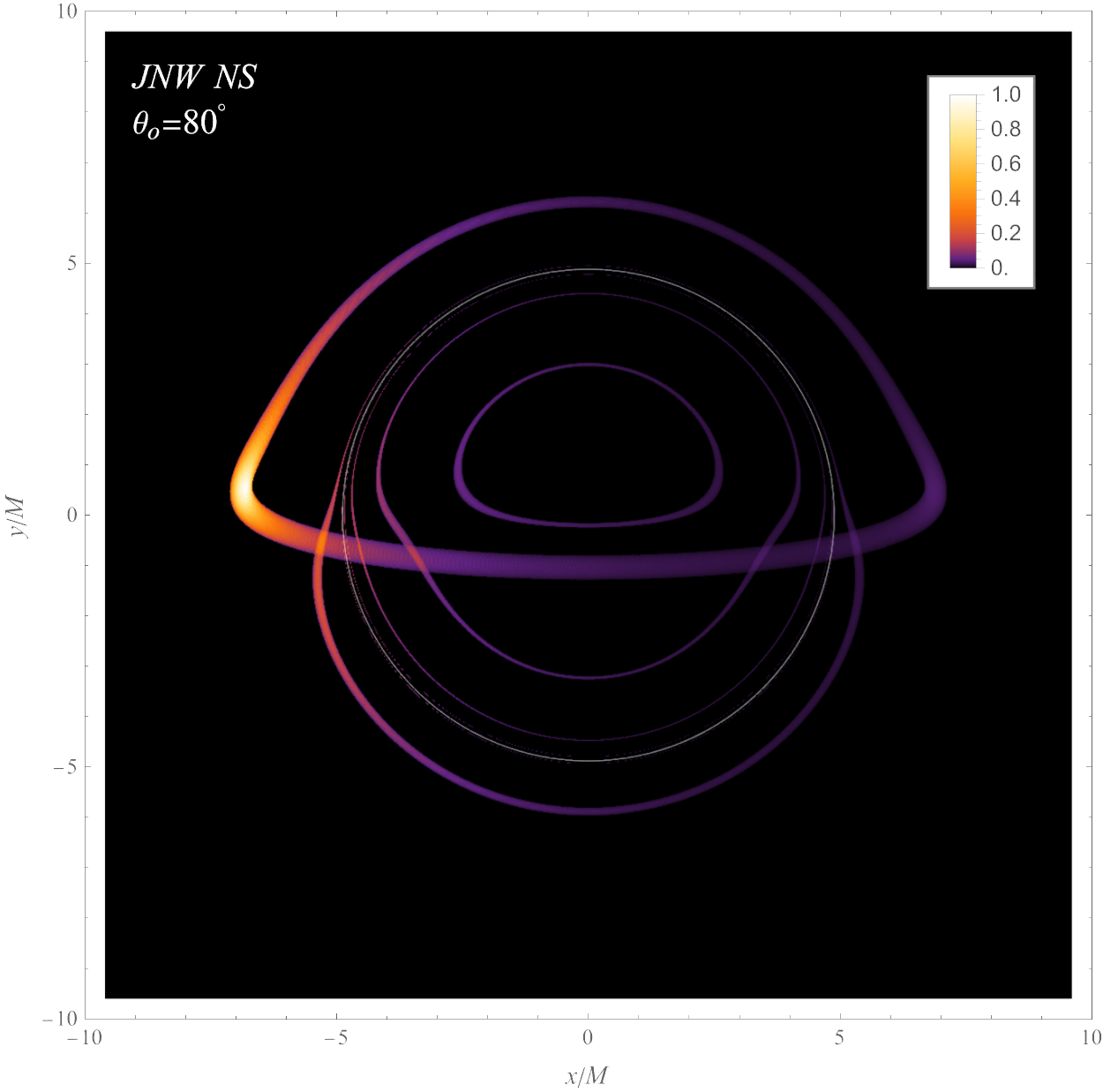}
\includegraphics[width=0.49\textwidth]{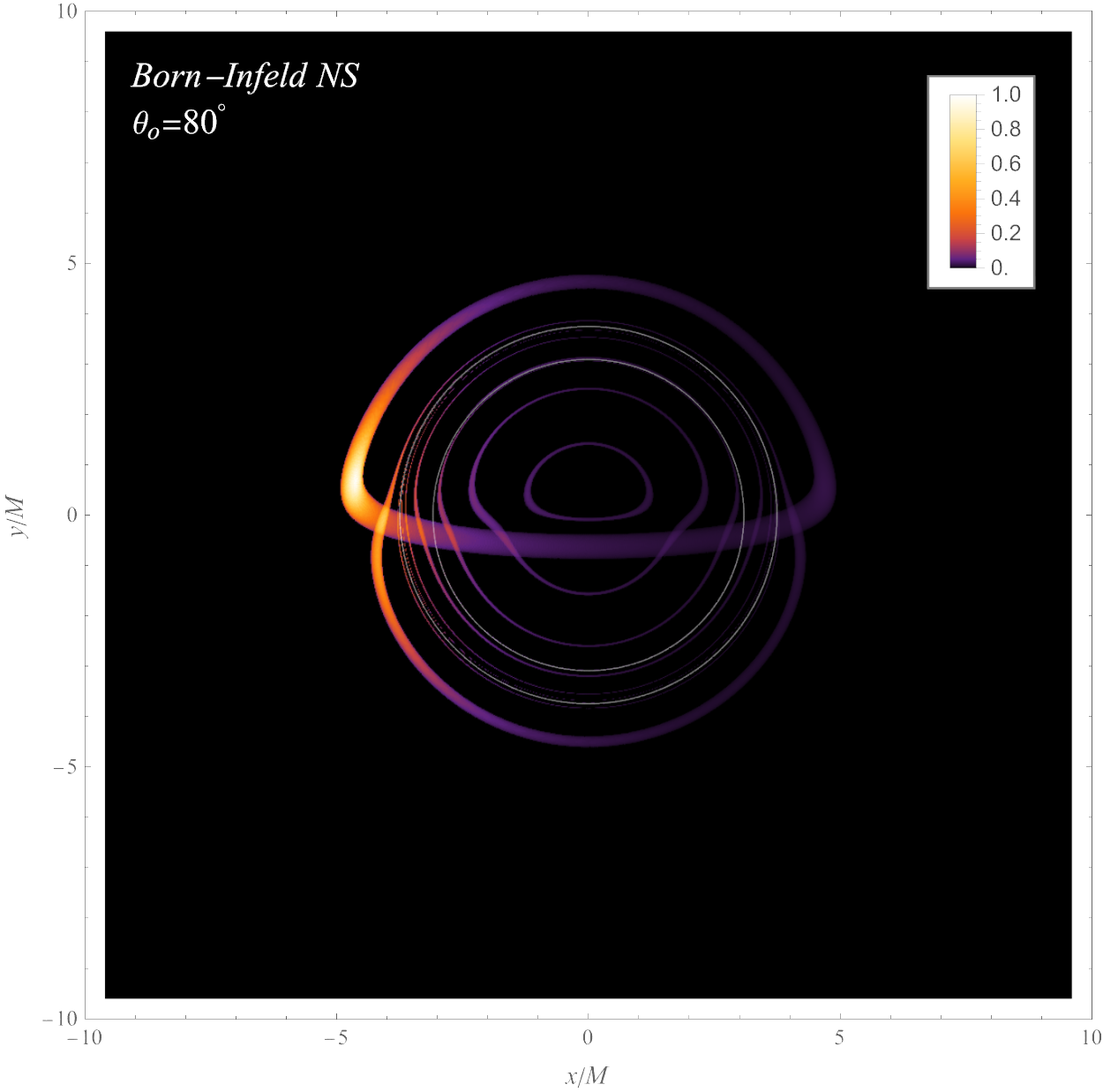}\caption{Time integrated images for a complete orbit of the hot spot, captured
from an observational inclination angle of $\theta_{o}=80^{\circ}$.
The white lines delineate the critical curves, shaped by light rays
that escape from the photon spheres. Intensity levels are normalized
to their maximum value. \textbf{Upper-Left Panel}: Schwarzschild black
hole. This image highlights the primary and secondary lensed image
tracks positioned beyond the critical curve, resulting from the $n=0^{>}$
and $1^{>}$ light rays emitted by the hot spot, respectively. \textbf{Upper-Right
Panel}: JNW singularity with $\gamma=0.9$. This image unveils two
image tracks outside the critical curve, alongside three additional
tracks within the critical curve. The latter tracks are produced by
$n=1^{<}$, $2^{<\text{ }}$and $3^{<}$ light rays traversing the
singularity. \textbf{Lower Panel}: Born-Infeld singularity with $a/M^{2}=1$,
$Q/M=1.05$ and $P/M=0$. The $n=3^{<>}$ and $4^{<>}$\ light rays,
engaged in orbits between the inner and outer photon spheres, create
two more image tracks situated amid the inner and outer critical curves.}
\label{TI-images-80}
\end{figure}

\begin{figure}[ptb]
\includegraphics[width=0.49\textwidth]{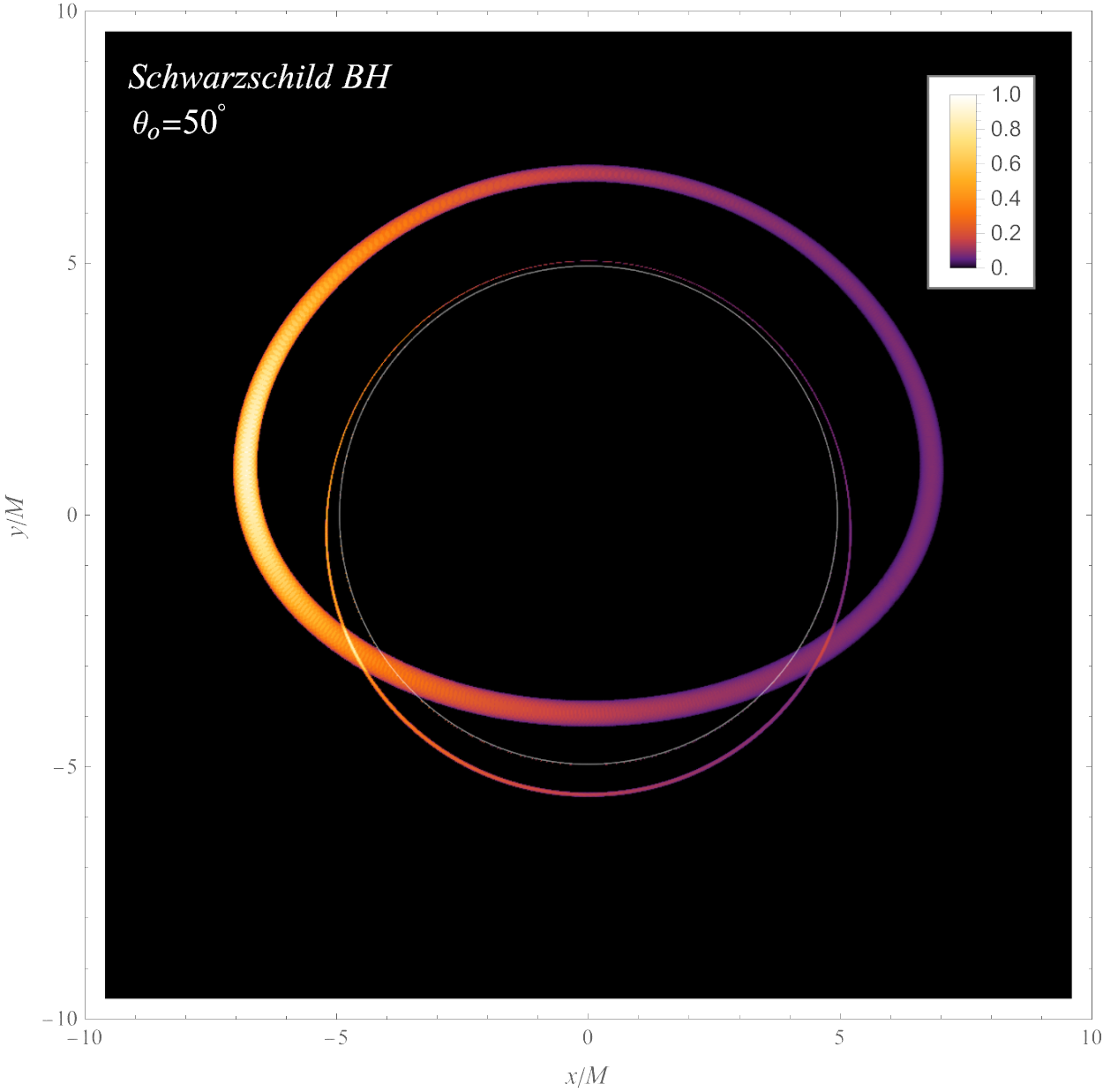}
\includegraphics[width=0.49\textwidth]{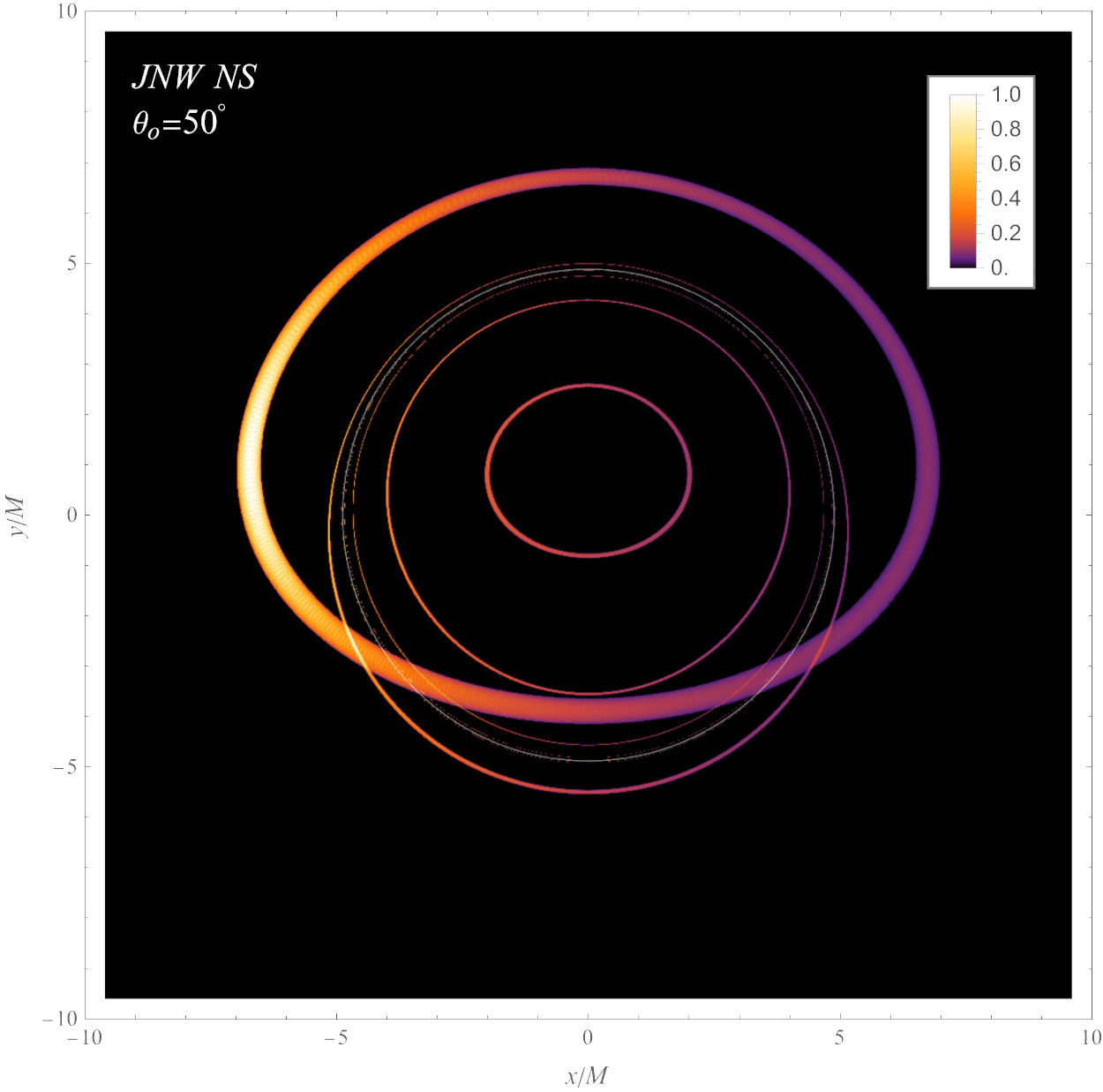}
\includegraphics[width=0.49\textwidth]{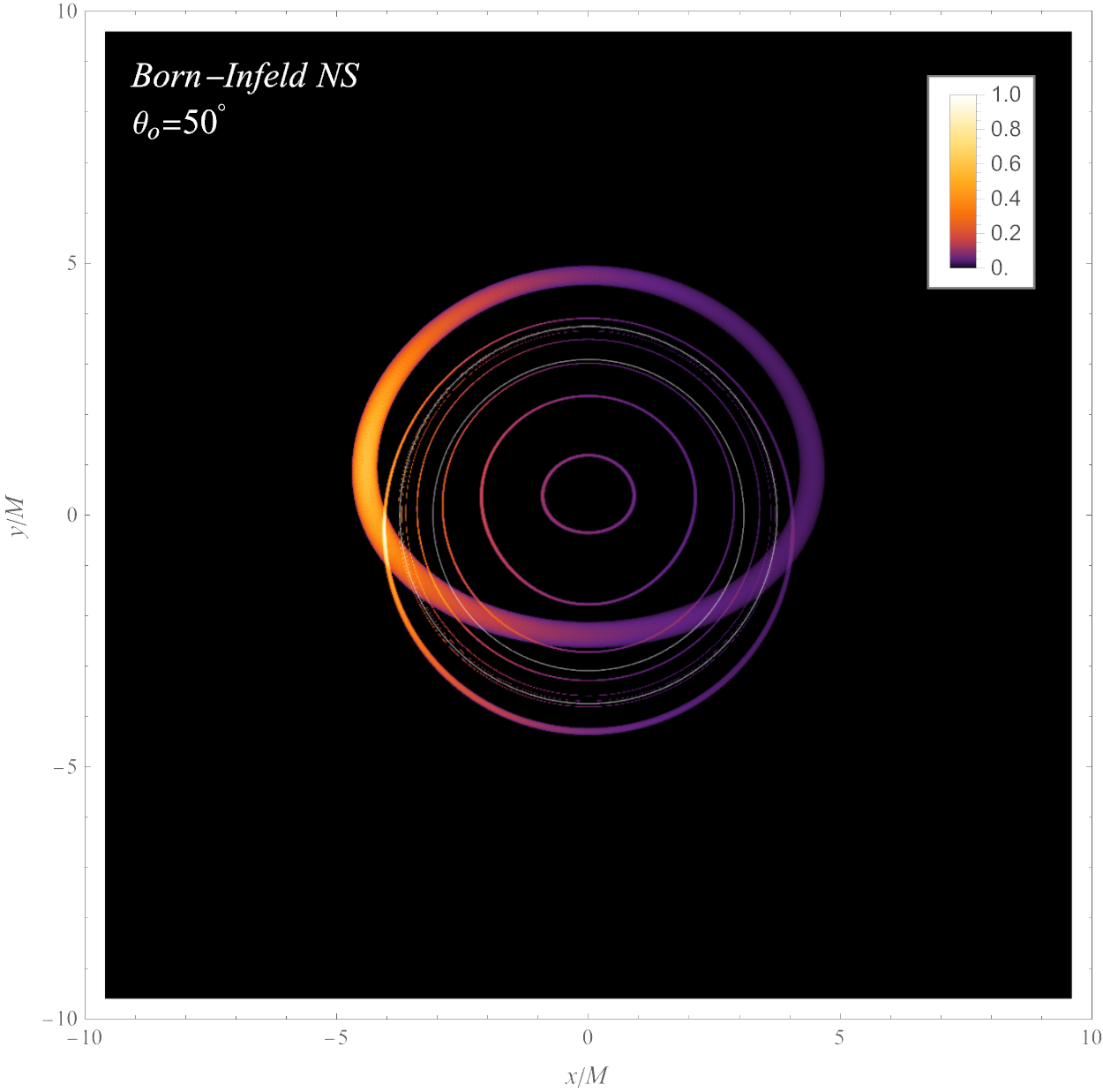}\caption{Time integrated images of the hot spot in the Schwarzschild black
hole (\textbf{Upper-Left Panel}), the JNW singularity (\textbf{Upper-Right
Panel}) and the Born-Infeld singularity (\textbf{Lower Panel}). The
observer inclination is $\theta_{o}=50^{\circ}$, with the central
object parameters being consistent with those in FIG. \ref{TI-images-80}.
Decreasing the inclination angle results in diminished brightness
asymmetry and the emergence of more circular image tracks.}
\label{TI-images-50}
\end{figure}

\begin{figure}[ptb]
\includegraphics[width=0.48\textwidth]{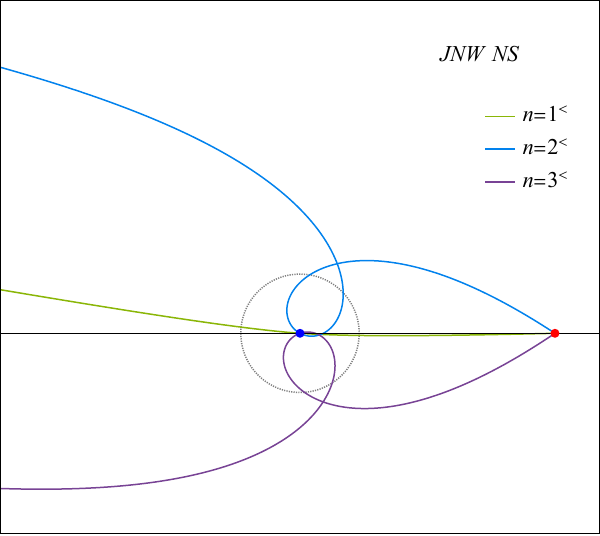} \includegraphics[width=0.48\textwidth]{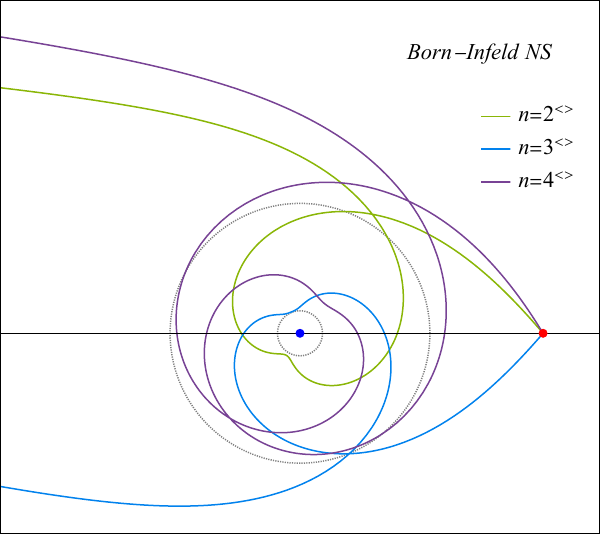}
\caption{Light rays connecting the hot spot to the observer with an observation
angle of $\theta_{o}=80^{\circ}$. The photon spheres are depicted
with dashed gray lines, while the singularity is marked by a blue
dot. \ \textbf{Left Panel}: Light rays responsible for generating
the image tracks inside the critical curve of the JNW singularity.
\textbf{Right Panel}: Light rays that produce the image tracks between
two critical curves of the Born-Infeld singularity. In $n$, the number
denotes the count of equatorial plane crossings by the light rays.
Furthermore, the $<$ and $<>$ correspond to light rays traversing
the singularities and orbiting between two photon spheres, respectively.}
\label{fig:geodesics}
\end{figure}

FIGs. \ref{TI-images-80} and \ref{TI-images-50} exhibit the time
integrated images for three distinct central objects, namely a Schwarzschild
black hole, a JNW singularity and a Born-Infeld singularity, as observed
from inclination angles of $\theta_{o}=80^{\circ}$ and $50^{\circ}$,
respectively. Here, we include observations of the hot spot orbiting
a Schwarzschild black hole to serve as both a run test for our code
and a benchmark for our analysis. As anticipated, the hot spot images
of the Schwarzschild black hole reveal two prominent bright image
tracks. Intriguingly, in contrast to the Schwarzschild black hole
case, the hot spot images in the JNW and Born-Infeld singularities
show additional tracks.

To understand the origin of these tracks, we present light rays of
interest connecting the hot spot at $\varphi=0$ to the observer at
$\theta_{o}=80^{\circ}$ in FIG. \ref{fig:geodesics}. We use a numerical
count, indicated as $n$, representing the number of times light rays
intersect the equatorial plane, as a means to characterize light rays
and consequently the resulting hot spot images. Furthermore, our previous
studies have demonstrated that light rays can pass through both the
JNW and Born-Infeld regularized singularities, thus giving rise to
a new set of images. Moreover, in the case of Born-Infeld singularities
featuring double photon spheres, photons are capable of orbiting the
singularities between the inner and outer photon spheres. Consequently,
we employ the superscripts $>$, $<>$ and $<$ to denote light rays
that travel outside the (outer) photon sphere, follow orbits between
the inner and outer photon spheres, and traverse the singularity inside
the (inner) photon sphere, respectively.

For an observer at an inclination angle of $\theta_{o}=80^{\circ}$,
FIG. \ref{TI-images-80} presents the time integrated images of the
hot spot. These images manifest a distinctive brightness asymmetry,
attributed to the Doppler effects. In the Schwarzschild black hole
case, the primary image with $n=0^{>}$ illustrates a closed semicircular
track, wherein its upper and lower segments depict the hot spot situated
behind and in front of the black hole. In contrast, the smaller and
dimmer track represents the secondary image with $n=1^{>}$. The scarcely
visible upper segment corresponds to the hot spot positioned in front
of the black hole, while the lower segment corresponds to the hot
spot positioned behind it. Furthermore, higher-order images exhibit
a markedly diminished luminosity and closely adhere to the critical
curve, which is formed by photons escaping from the photon sphere.

In addition to the two previously mentioned image tracks, the time
integrated image of the hot spot in the JNW singularity, displayed
in the upper-right panel of FIG. \ref{TI-images-80}, reveals the
presence of three additional tracks. These three tracks are formed
by light rays that traverse the singularity, positioning them within
the critical curve. Specifically, moving from the innermost to the
outermost region, these image tracks arise from the $n=1^{<}$, $2^{<}$
and $3^{<}$ light rays, as visually depicted in the left panel of
FIG. \ref{fig:geodesics}. Furthermore, the upper and lower segments
of the $n=1^{<}$ and $3^{<}$ tracks correspond, respectively, to
the images of the hot spot located in front of and behind the singularity.
Additionally, the $n=2^{<}$ track features upper and lower segments
corresponding, respectively, to the images of the hot spot located
behind and in front of the singularity.

Due to the presence of double photon spheres, the hot spot image in
the Born-Infeld singularity, shown in the lower panel of FIG. \ref{TI-images-80},
exhibits two critical curves. Analogously, the image tracks corresponding
to $n=0^{>}$ and $1^{>}$ are observed outside the outermost critical
curve, while the image tracks linked to $n=1^{<}$, $2^{<\text{ }}$and
$3^{<}$ are discernible within the innermost critical curve. However,
light rays engaged in orbits between the inner and outer photon spheres
contribute additional image tracks located between the two critical
curves. In particular, the $n=3^{<>}$ and $4^{<>}$ light rays form
two visible image tracks between the inner and outer critical curves.
Yet, due to strong gravitational bending amid the two photon spheres,
the $n=2^{<>}$ light rays produce a faint crescent shape at the summit
of the inner critical curve. The $n=2^{<>}$, $3^{<>}$ and $4^{<>}$
light rays are depicted in the right panel of FIG. \ref{fig:geodesics}.

FIG. \ref{TI-images-50} depicts the hot spot images obtained at an
observation inclination of $\theta_{o}=50^{\circ}$, revealing a certain
similarity between this case and the one with $\theta_{o}=80^{\circ}$.
Despite this similarity, it becomes evident that the observer, positioned
at a lower inclination angle, witnesses a diminished level of brightness
asymmetry, while the image tracks adopt a more circular form. Furthermore,
our findings indicate that, in the Born-Infeld singularity, the $n=2^{<>}$
light rays emitted from the hot spot fail to reach the observer at
$\theta_{o}=50^{\circ}$.

\subsection{Temporal Fluxes and Centroids}

\begin{figure}[ptb]
\includegraphics[width=0.33\textwidth]{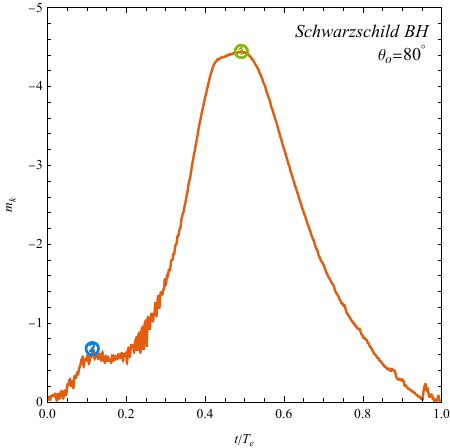}\includegraphics[width=0.33\textwidth]{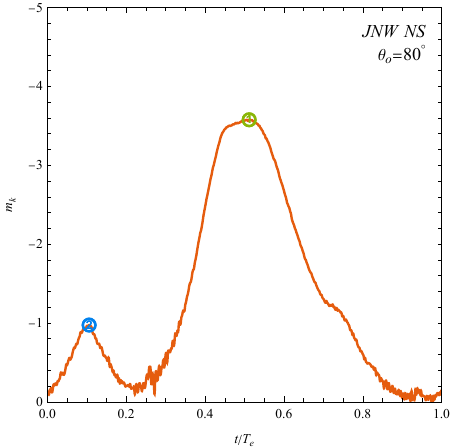}\includegraphics[width=0.33\textwidth]{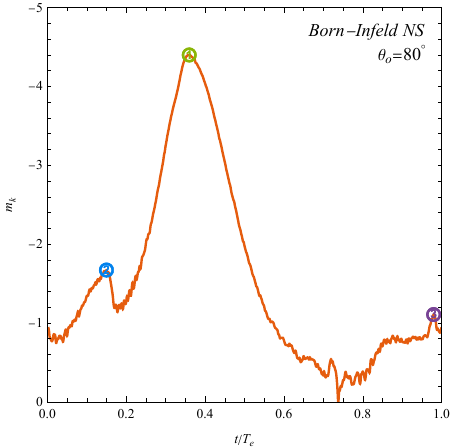} 

\includegraphics[width=0.33\textwidth]{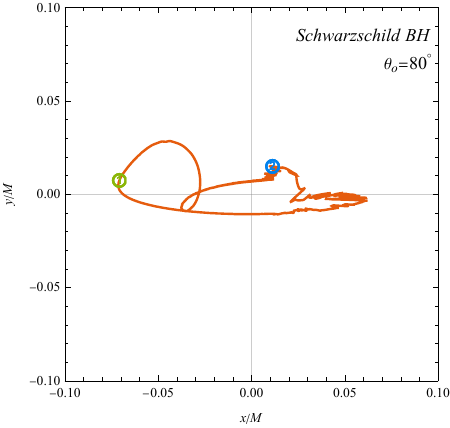}\includegraphics[width=0.33\textwidth]{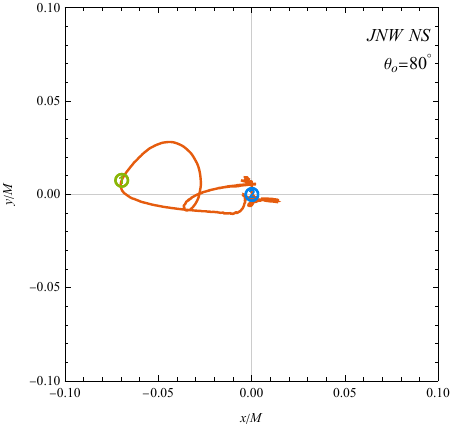}\includegraphics[width=0.33\textwidth]{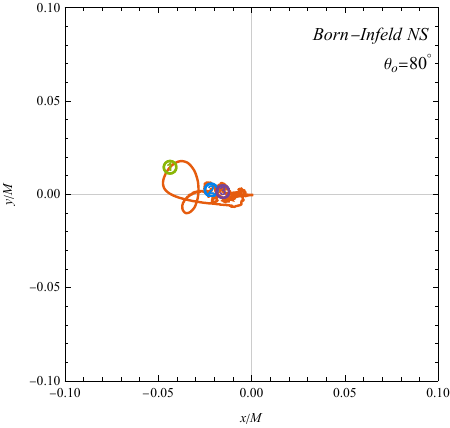}\caption{Temporal magnitudes $m_{k}$ (\textbf{Upper Row}) and centroids $c_{k}$
(\textbf{Lower Row}) as a function of $t/T_{e}$ for the Schwarzschild
black hole (\textbf{Left Column}), the JNW singularity (\textbf{Middle
Column}) and the Born-Infeld singularity (\textbf{Right Column}).
The inclination is $\theta_{o}=80^{\circ}$. The highest and second-highest
peaks of the temporal magnitude are indicated by \textcolor{darkpastelgreen}{\textcircled{1}}
and \textcolor{denim}{\textcircled{2}}, respectively. The presence
of additional higher-order images results in the JNW and Born-Infeld
cases having a more pronounced second-highest peak compared to the
Schwarzschild black hole case. Similarly, a third-highest peak, designated
as \textcolor{electricviolet}{\textcircled{3}}, emerges in the
Born-Infeld singularity. The centroids of the flux at these peaks
are identified with corresponding numbers.}
\label{fig:mkck-80}
\end{figure}

\begin{figure}[ptb]
\includegraphics[width=0.333\textwidth]{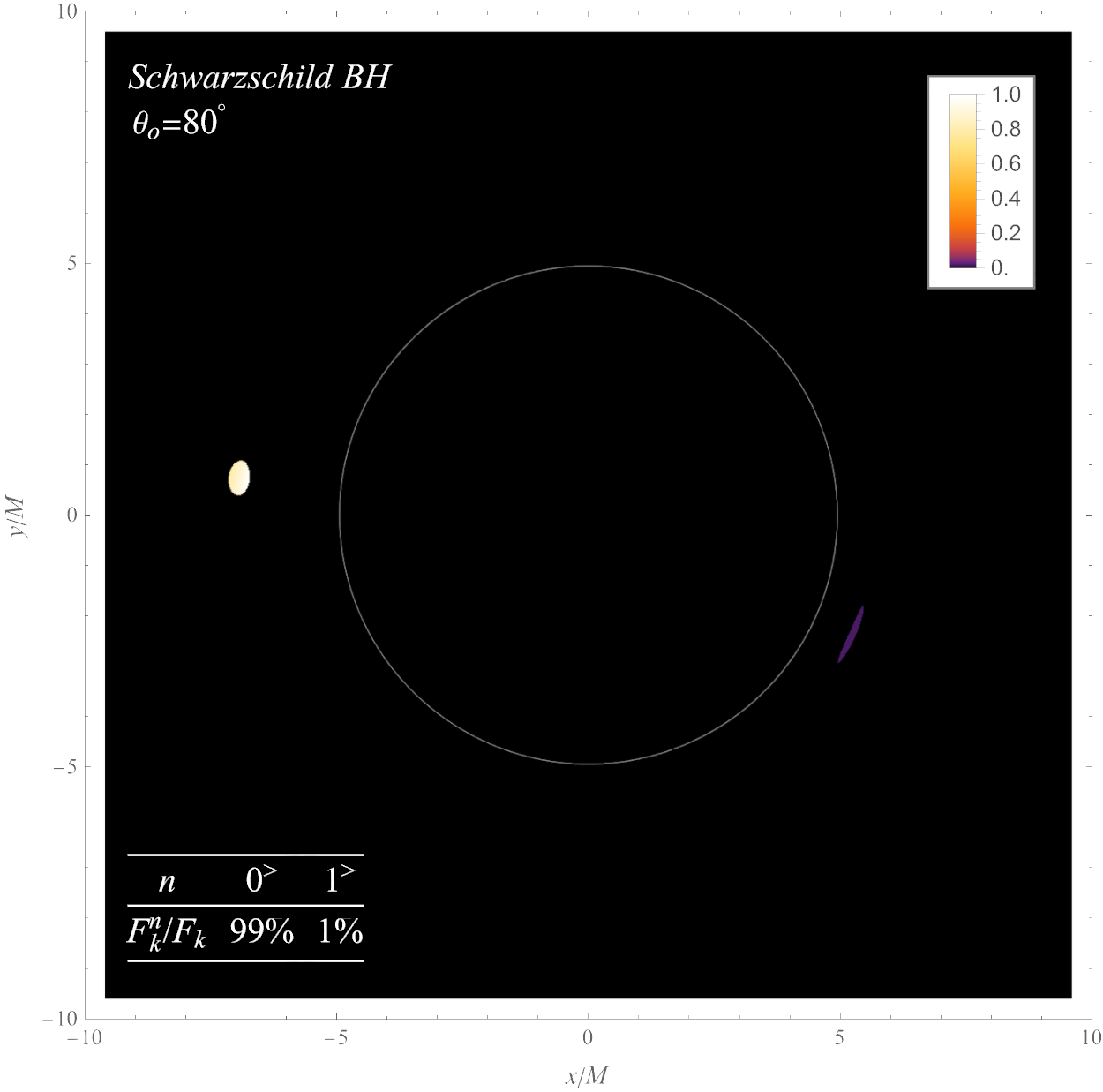}\includegraphics[width=0.333\textwidth]{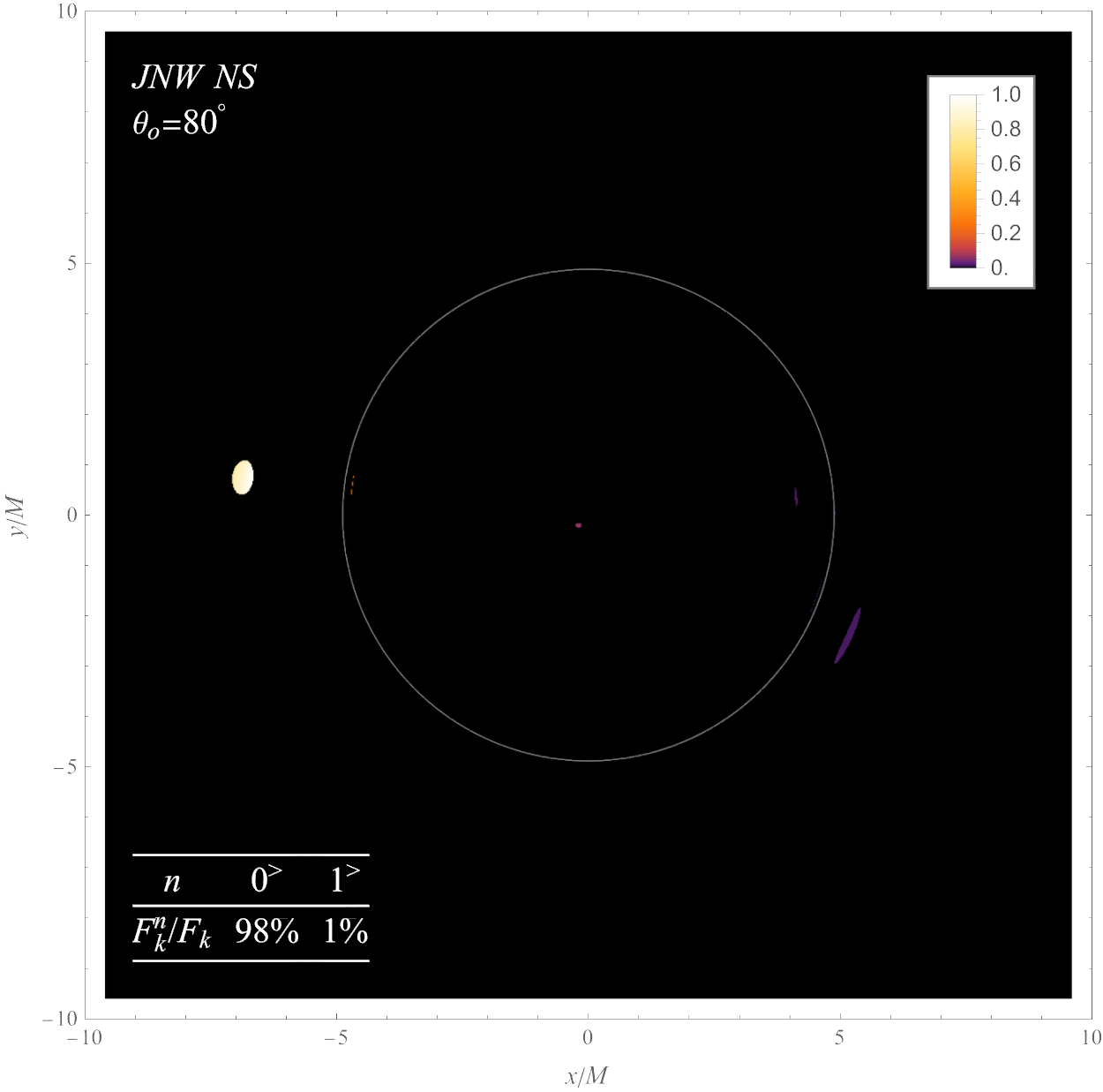}\includegraphics[width=0.333\textwidth]{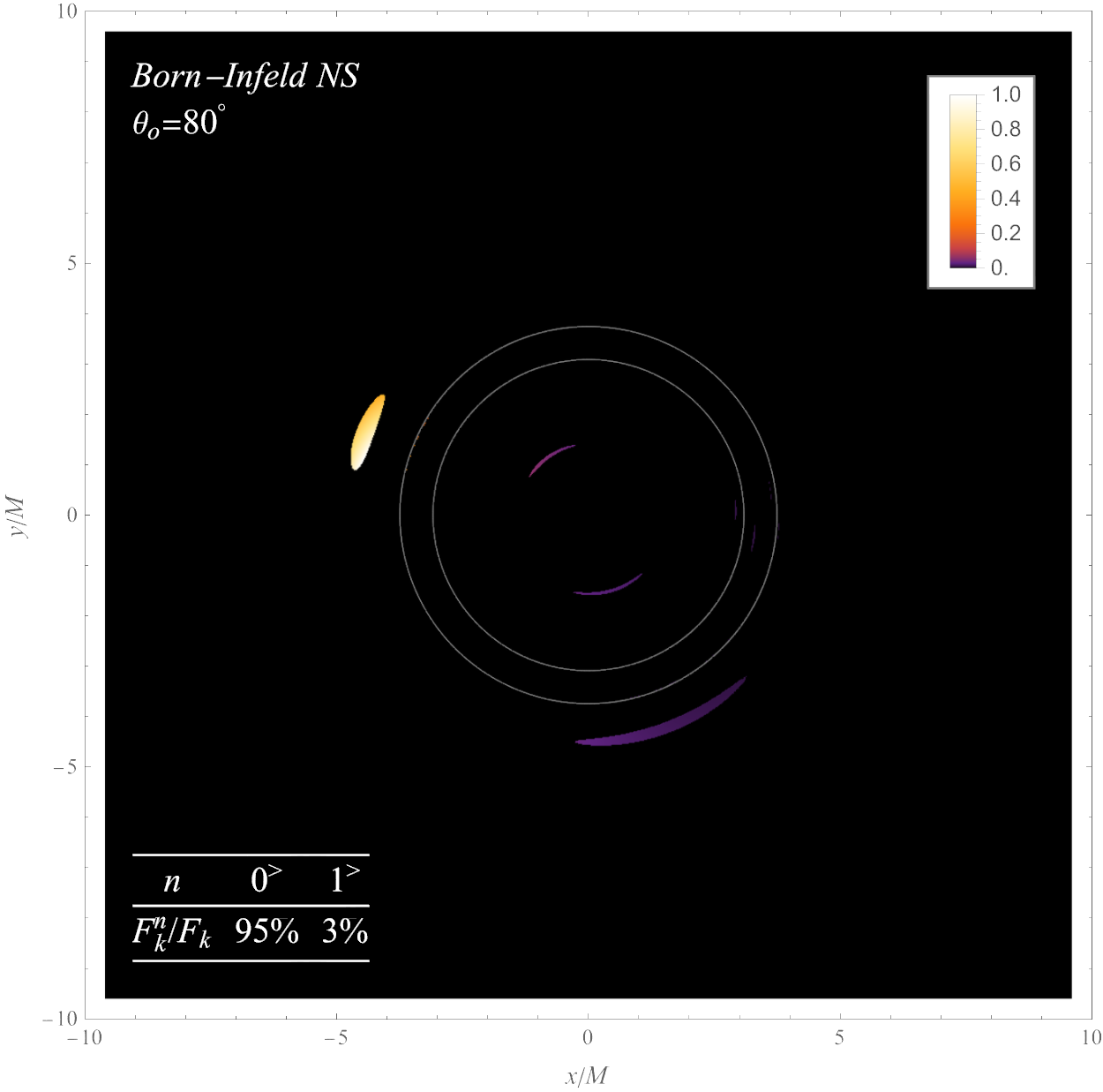} 

\includegraphics[width=0.333\textwidth]{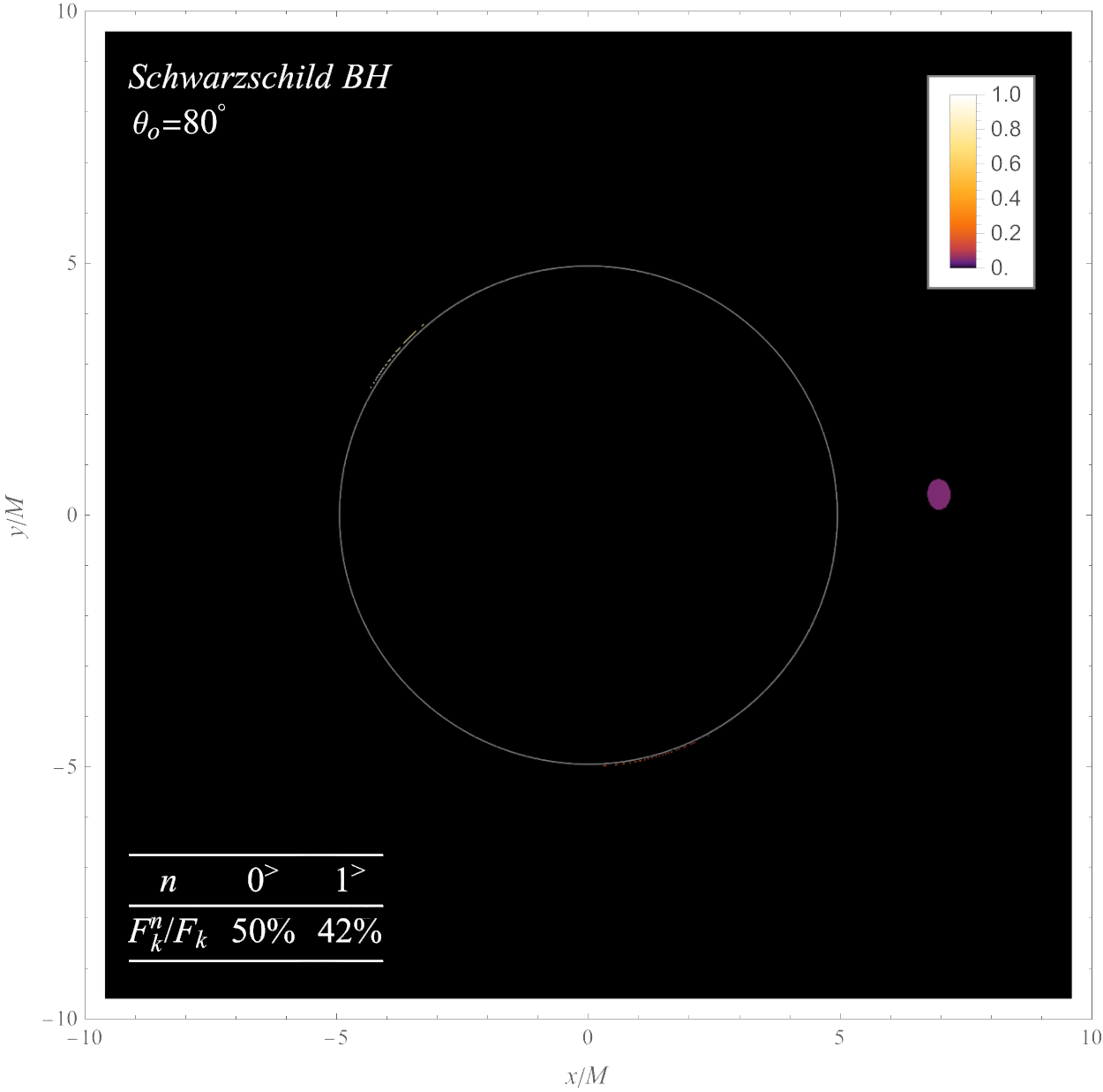}\includegraphics[width=0.333\textwidth]{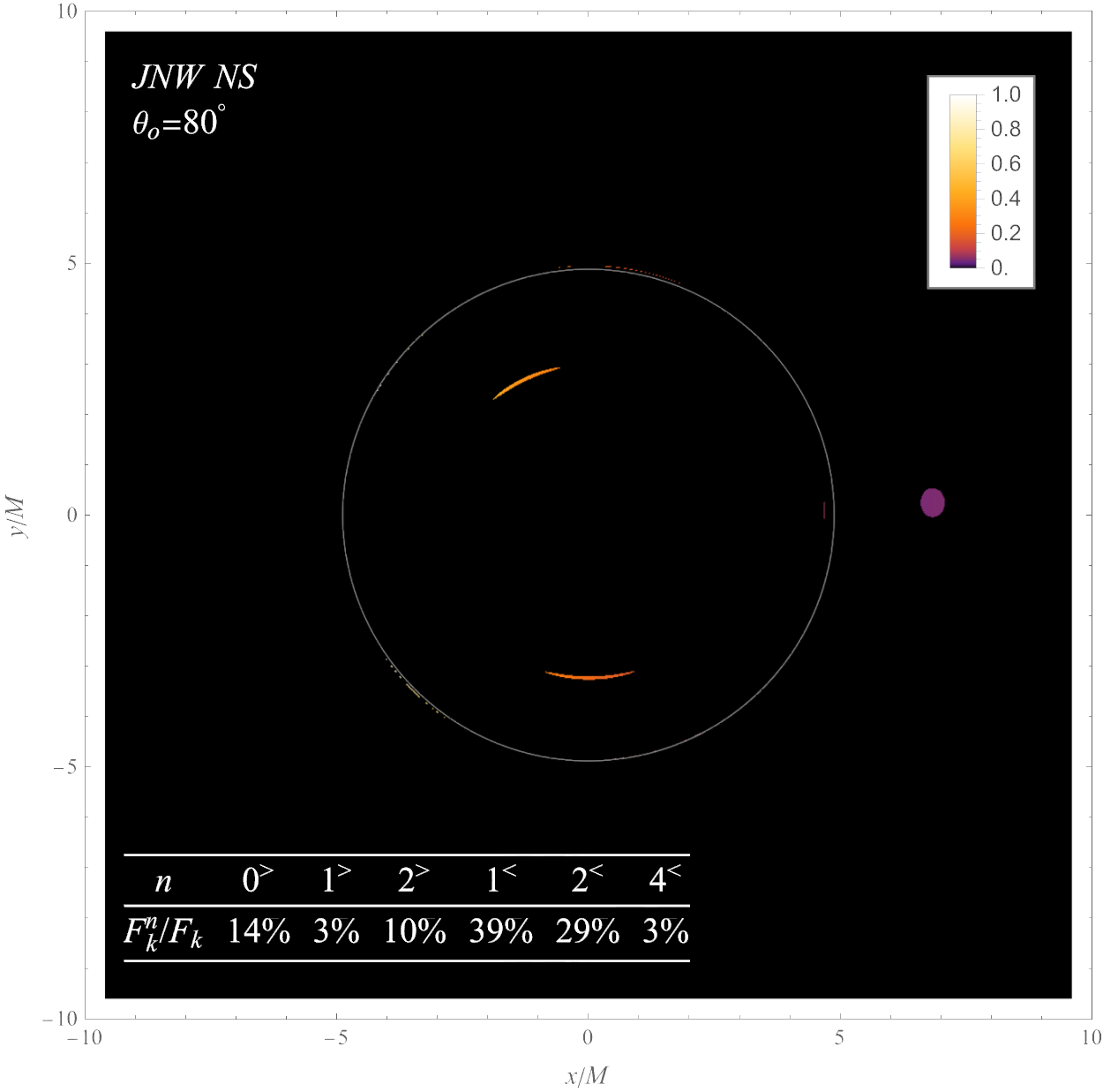}\includegraphics[width=0.333\textwidth]{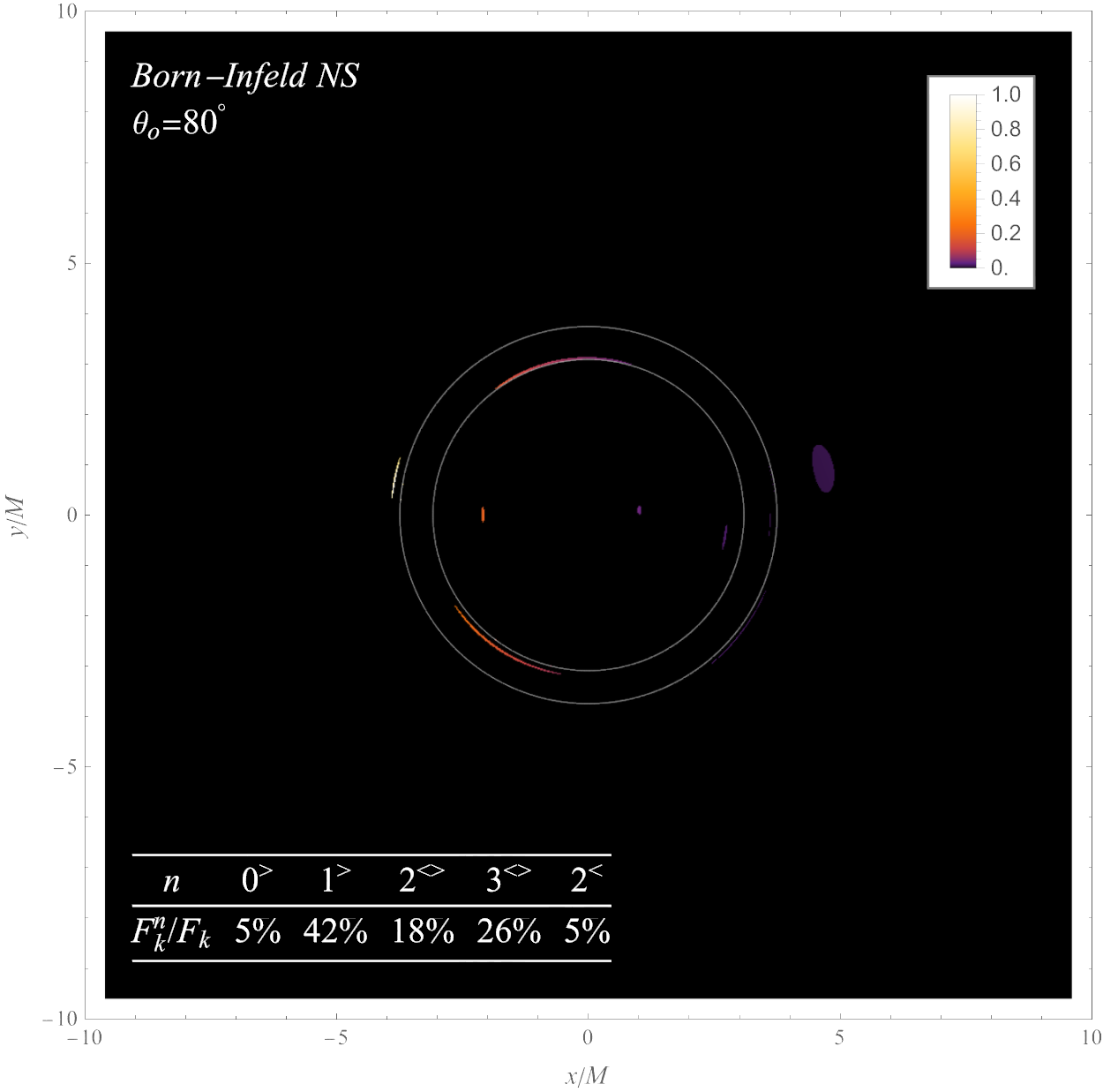} 

\hfill{} \includegraphics[width=0.33\textwidth]{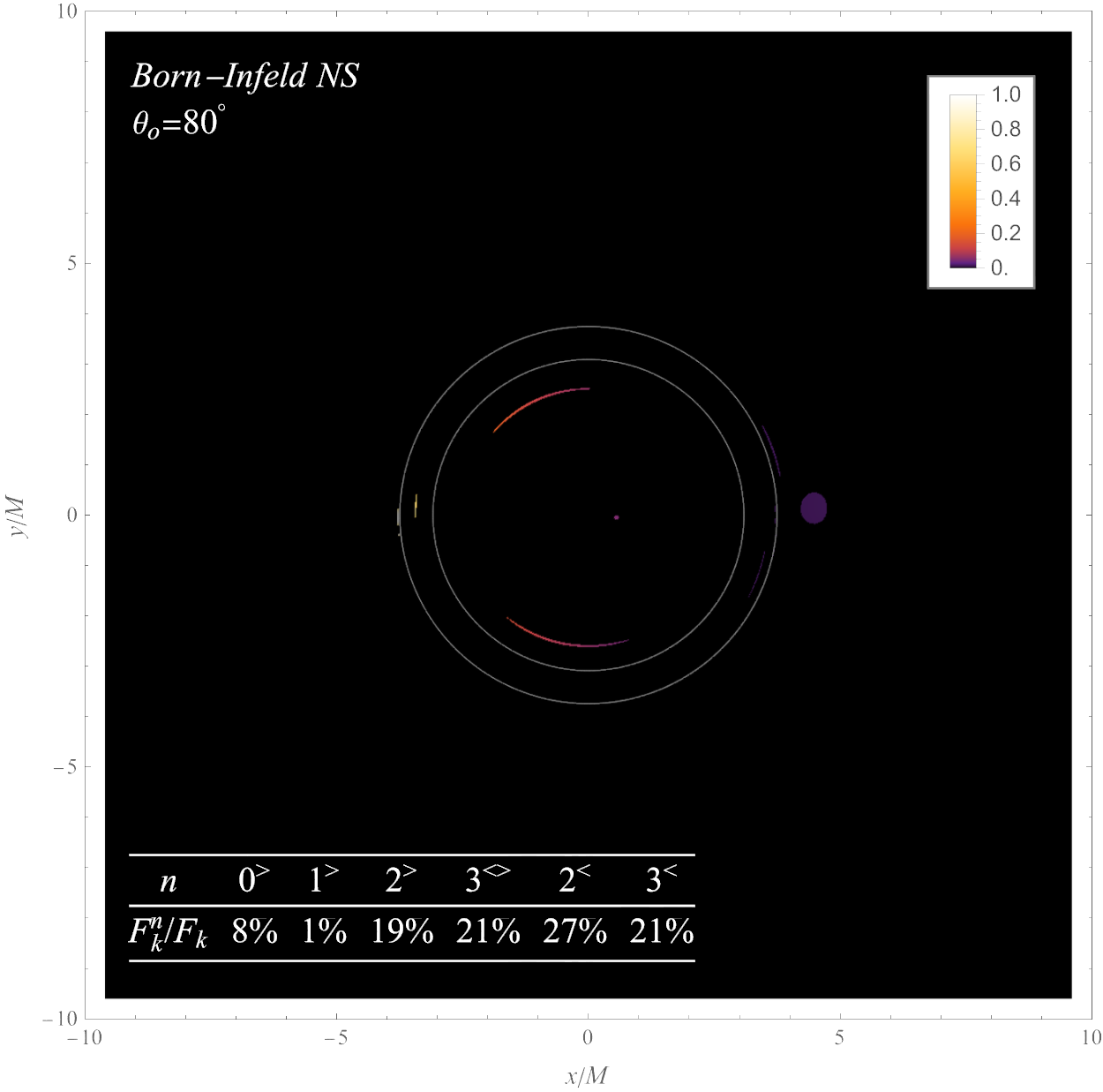}
\label{mkck-80}\caption{Snapshots for the Schwarzschild black hole (\textbf{Left Column}),
the JNW singularity (\textbf{Middle Column}) and the Born-Infeld singularity
(\textbf{Right Column}) when the temporal magnitude reaches its maximum
value. The top, middle and bottom rows present the snapshots at the
highest, second-highest and third-highest peaks, respectively. The
contribution from the $n$th-order image to the total flux is quantified
by $F_{k}^{n}/F_{k}$, where $F_{k}^{n}$ represents the temporal
flux of the $n$th-order image at $t=t_{k}$.}
\label{fig:snapshots}
\end{figure}

FIG. \ref{fig:mkck-80} illustrates the temporal magnitudes $m_{k}$
and centroids $c_{k}$ observed by an observer positioned at an inclination
angle of $\theta_{o}=80^{\circ}$ for the Schwarzschild black hole,
the JNW singularity and the Born-Infeld singularity. In the cases
of the Schwarzschild black hole and JNW singularity, two distinct
peaks in the temporal magnitude are evident, denoted by \textcolor{darkpastelgreen}{\textcircled{1}}
and \textcolor{denim}{\textcircled{2}}. Meanwhile, the Born-Infeld
singularity displays an additional third peak, marked by \textcolor{electricviolet}{\textcircled{3}}.
The snapshots corresponding to the highest peak of the temporal magnitudes
are displayed in the upper row of FIG. \ref{fig:snapshots}, along
with the contributions from relevant images with different values
of $n$, as listed in the accompanying tables. It is evident that
the flux at the highest peak is predominantly governed by the primary
images with $n=0^{>}$, generated by the hot spot positioned near
the leftmost portion of the orbit. This observation aligns with expectations,
as the hot spot moves closer to the observer on the left side of the
field of view, leading to a pronounced increase in the observed light
frequency due to the Doppler effect.

Of particular interest is the second-highest peak, indicated by \textcolor{denim}{\textcircled{2}},
which is notably more prominent in the JNW and Born-Infeld singularities
compared to the Schwarzschild black hole case. Furthermore, the corresponding
snapshots are depicted in the middle row of FIG. \ref{fig:snapshots},
revealing a shift away from exclusive dominance by primary images
in terms of flux. In fact, primary images experience a phase of reduced
flux as the hot spot moves away from the observer, leading to a decrease
in the observed frequency. If other images achieve their peak flux
values, they can produce localized total flux peaks. In the Schwarzschild
black hole, while the $n=1^{>}$ image substantially contributes to
the total flux, the primary image still contributes $50\%$, resulting
in an insignificant peak. Conversely, in the context of the JNW singularity,
$n=1^{<}$ and $2^{<}$ images emerge as two crucial contributors
to the total flux, leading to a notably pronounced local peak. Similarly,
in the case of the Born-Infeld singularity, the presence of the $n=2^{<>}$,
$3^{<>}$ and $2^{<}$ images contributes to a noticeable peak in
$m_{k}$. Furthermore, the snapshot in the bottom row of FIG. \ref{fig:snapshots}
demonstrates that the third peak in $m_{k}$, identified by \textcolor{electricviolet}{\textcircled{3}}
in the upper-right panel of FIG. \ref{fig:mkck-80}, emerges from
images within the inner critical curve.

The absence of higher-order images would lead the temporal centroid
to align with the center of the primary image. Nonetheless, when Doppler
effect-induced flux reduction affects the primary image, the presence
of higher-order images can markedly displace the centroid away from
the center of the primary image's orbit. In comparison with the Schwarzschild
black hole case, extra higher-order images tend to displace the centroid
more significantly to the left in the image plane for both the JNW
and Born-Infeld singularities. Additionally, due to contributions
from higher-order images in close proximity to critical curves, numerical
noise becomes evident in the low flux region, affecting the temporal
magnitudes and centroids.

\begin{figure}[ptb]
\includegraphics[width=0.33\textwidth]{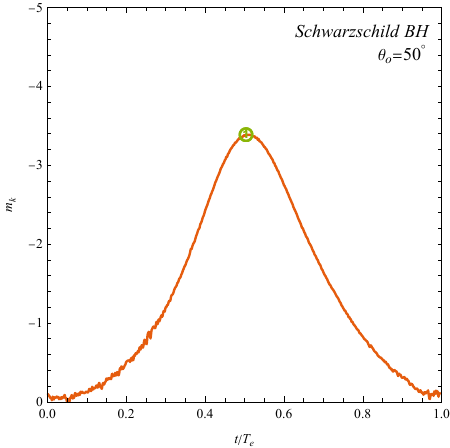}\includegraphics[width=0.33\textwidth]{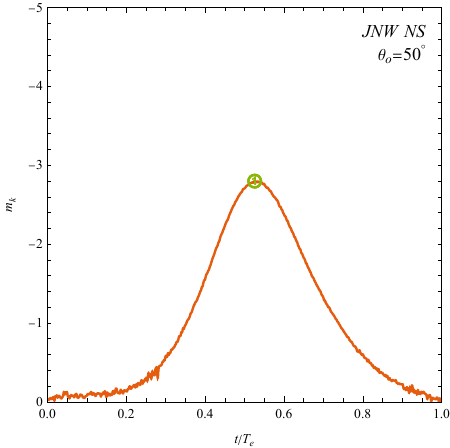}\includegraphics[width=0.33\textwidth]{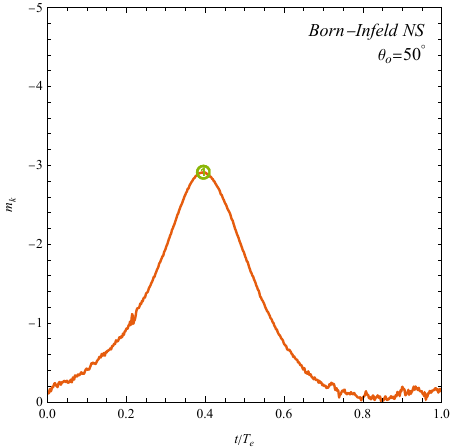} 

\includegraphics[width=0.33\textwidth]{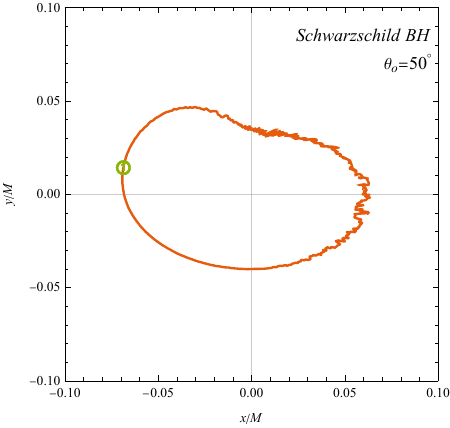}\includegraphics[width=0.33\textwidth]{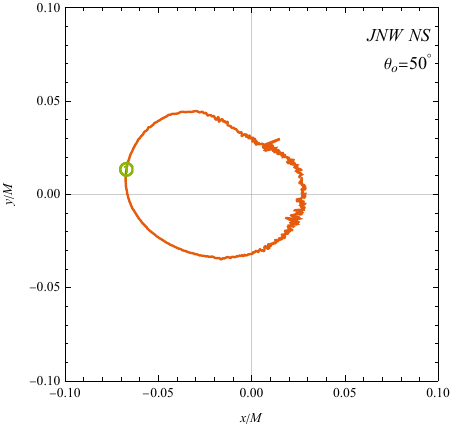}\includegraphics[width=0.33\textwidth]{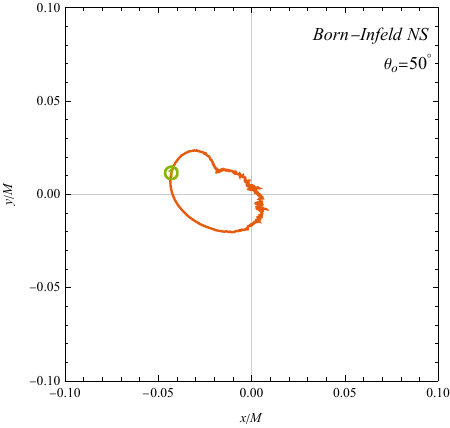}\caption{Temporal magnitudes $m_{k}$ (\textbf{Upper Row}) and centroids $c_{k}$
(\textbf{Lower Row}) as a function of $t/T_{e}$ for the Schwarzschild
black hole (\textbf{Left Column}), the JNW singularity (\textbf{Middle
Column}) and the Born-Infeld singularity (\textbf{Right Column}).
The inclination is $\theta_{o}=50^{\circ}$. Given the diminished
impact of the Doppler effect at low inclinations, the temporal magnitudes
exhibit a single peak across all cases.}
\label{mkck-50}
\end{figure}

The temporal magnitudes and centroids for an inclination angles of
$\theta_{o}=50^{\circ}$ are presented in FIG. \ref{mkck-50}. In
contrast to the $\theta_{o}=80^{\circ}$ inclination, a sole peak
is evident in the temporal magnitudes for $\theta_{o}=50^{\circ}$.
This dissimilarity emerges due to the reduced influence of the Doppler
effect at the lower inclination. As a result, the flux becomes less
dependent on the frequency, allowing the primary image to dominate
most of the time in the contribution to the total flux. Accordingly,
the effect of higher-order images on centroids is reduced, resulting
in a less intricate trajectory for the centroid's orbit.

\section{Conclusions}

\label{sec:CONCLUSIONS}

This paper investigated observations of hot spots in the JNW and Born-Infeld
naked singularities as they move along the ISCOs. Intriguingly, in
these spacetimes, photons have been observed to reach the singularity
in a finite coordinate time once they enter the photon spheres \cite{Chen:2023trn,Chen:2023uuy}.
Furthermore, when the singularity is regularized, these photons are
capable of traversing the regularized singularity, thereby generating
new images of hot spots positioned within the critical curves. Particularly,
in contrast to Schwarzschild black holes, JNW and Born-Infeld singularities
exhibit numerous additional image tracks in the time integrated images
that capture a full orbit of hot spots. Consequently, when observed
at low inclinations, these extra images result in a more pronounced
second-highest peak in the temporal magnitudes in the JNW and Born-Infeld
singularities. Additionally, a third peak can arise in the Born-Infeld
singularity spacetime.

As discussed in \cite{Chen:2023uuy}, optical appearances of hot spots
depend on the regularization schemes of the singularities. In cases
where the regularized singularity spacetime models a traversable wormhole
\cite{Pal:2022cxb}, hot spots in our universe exhibit appearances
akin to those of black holes. Conversely, for hot spots situated in
another universe, only images positioned within the critical curve
are observable. The emergence of the next-generation Very Long Baseline
Interferometry offers promising prospects for utilizing our discoveries
as a tool to investigate the nature of naked singularities.
\begin{acknowledgments}
We are grateful to Qingyu Gan and Xin Jiang for useful discussions
and valuable comments. This work is supported in part by NSFC (Grant
No. 12105191, 12275183, 12275184 and 11875196). Houwen Wu is supported
by the International Visiting Program for Excellent Young Scholars
of Sichuan University. 
\end{acknowledgments}

 \bibliographystyle{unsrturl}
\bibliography{Ref}

\end{document}